\documentclass{article}
\usepackage{spconf,amsmath,graphicx,hyperref}
\usepackage{cite}
\usepackage{amsmath,amssymb,amsfonts}
\usepackage{graphicx}
\usepackage{booktabs}
\usepackage{balance}
\graphicspath{{Figures/}}

\title{Anchor-Free Range-Free 3D IoT Localization in Buried Environments with Connectivity-Informed Power Selection and Temporal Refinement}
\name{Peyman Moeini, Mark Coates\thanks{This work was supported in part by the McGill University Vadasz Scholars Program.}}
\address{McGil University \\
Department of Electrical and Computer Engineering}
\begin{document}
%
\maketitle
\begin{abstract}
In underground and surface mines, wireless Internet-of-Things (IoT) tags deployed with ore or muck loads can support material and grade tracking after burial, when conventional ranging and surveyed localization infrastructure may be unavailable. Building on the three-dimensional Protraction Technique (3D-PT), this paper introduces 3D-PT with Power Selection (3D-PT-PS), which selects informative transmit-power levels from binary pilot measurements, and 3D-PT-PS with Temporal Refinement (3D-PT-PST), which incorporates recorded load-deposition times as weak, non-metric cues. Both methods are anchor-free and range-free and require no RSSI, measured ranges, surveyed anchors, pile dimensions, or material parameters. Monte Carlo simulations evaluate different pile sizes and tag densities. In a $5000$-t crusher-pocket scenario with 125 tags, 3D-PT-PST achieves $6.12\%$ mean NRMSE and $2.20$~m 3D RMSE, reducing mean NRMSE by $13.5\%$ relative to 3D-PT-PS and by $43.6\%$ and $69.1\%$ relative to calibrated MDS-MAP(P) and CATL, respectively. These results show that connectivity and timing provide complementary information.
\end{abstract}

\section{Introduction}
\label{sec:introduction}

In mining, blast-fragmented material (muck) from different headings or blast locations can vary in ore grade, i.e., the concentration of valuable mineral in the material, and commonly passes through multiple handling and stockpiling stages before reaching the processing plant (mill) \cite{marinin2021assessing}. As truckloads are combined, buried, mixed, or rehandled, load provenance and grade become difficult to preserve \cite{vanduijvenbode2020performance}. Low-power IoT tags deployed with selected loads can preserve load identity and, when localized within the pile, help track where material of different provenance and expected grade resides \cite{young2021modelling}. Radio propagation through dense, heterogeneous muck can experience strong and spatially variable attenuation \cite{theissen2023usecase}, while settling, mixing, and rehandling can alter the tags' positions after deposition \cite{both2021applied}. Surveyed anchors and RSSI-, time-, or angle-based ranging may therefore be unavailable or unreliable \cite{geok2021review}.

The Protraction Technique (PT) was proposed for two-dimensional localization \cite{moeini2026protraction} and later extended to the three-dimensional Protraction Technique (3D-PT) \cite{moeini2026protraction3d}. 3D-PT converts multi-level binary reachability into reliability-weighted ordinal relations for relative 3D reconstruction \cite{moeini2026protraction3d}. Our work builds on 3D-PT for low-power IoT tags embedded in bulk material, where tags may be deposited sequentially with successive truck loads.

We introduce \emph{3D-PT with Power Selection (3D-PT-PS)}, which augments 3D-PT with connectivity-informed selection of transmit-power levels from binary pilot graphs rather than relying on a predefined power ladder. We further introduce \emph{3D-PT-PS with Temporal Refinement (3D-PT-PST)}, which uses the recorded time at which each tagged load is deposited into a monitored bulk-material region (e.g., a stockpile) as a weak, non-metric cue. We evaluate both methods against Multidimensional Scaling using patches of relative maps (MDS-MAP(P)) and a reproduction of Connectivity-based and Anchor-free Three-dimensional Localization (CATL) \cite{shang2004improved,tan2013catl}. Relative to 3D-PT, the proposed methodological extensions are:
(i) connectivity-informed binary power-ladder selection using pilot graph statistics; and
(ii) temporal refinement along an internally inferred deposition axis.
We additionally conduct a fixed-budget $K$-ablation to quantify the benefit of additional reachability levels in dense materials.

\section{System Model and Problem Formulation}
\label{sec:system_model}

Consider $n$ low-power IoT tags embedded within a 3D bulk-material
region. Indices $i,j,\ell\in\{1,\ldots,n\}$ denote tags and
$k\in\{1,\ldots,K\}$ indexes transmit-power levels. Tag $i$ has unknown
position $\mathbf{x}_i\in\mathbb{R}^{3}$, with
$\mathbf{X}=[\mathbf{x}_1,\ldots,\mathbf{x}_n]^{\mathsf T}
\in\mathbb{R}^{n\times3}$.

The network is anchor-free and range-free: no localization method is
provided with true tag coordinates or measured inter-tag distances.

Each tag communicates using ordered transmit-power levels
$\mathcal{P}=\{p_1,\ldots,p_K\}$, with $p_1<\cdots<p_K$.
For ordered pair $(i,j)$, $a_{ij}^{(k)}=1$ if the detected packet
fraction from tag $i$ to tag $j$ at $p_k$ meets the binary detection
threshold, and $0$ otherwise. Thus,
$\mathbf{A}^{(k)}=[a_{ij}^{(k)}]$ is the directed binary detection graph,
with a reciprocal edge between $i$ and $j$ when
$a_{ij}^{(k)}a_{ji}^{(k)}=1$.

Following 3D-PT, 3D-PT-PS converts the ordered binary observations into a pairwise reachability score $s_{ij}$ and reliability coefficient $\kappa_{ij}\in[0,1]$ \cite{moeini2026protraction3d}. These quantities induce local and global ordinal relations such as
\begin{equation}
d_{ij}<d_{i\ell},
\qquad
d_{ij}<d_{uv},
\qquad
d_{ij}=\|\mathbf{x}_i-\mathbf{x}_j\|_2.
\end{equation}
Here, $(u,v)$ denotes a second tag pair, so $d_{ij}<d_{uv}$ compares
the separations of two tag pairs.

The goal is to recover a relative configuration $\widehat{\mathbf{X}}$ consistent with ordinal relations derived from binary connectivity and, for 3D-PT-PST, temporal deposition information. Without a surveyed spatial reference, the map is determined only up to translation, rotation or reflection, and global scale. Ground truth is used only for post-localization alignment and evaluation.

\section{3D-PT with Power Selection (3D-PT-PS) and Temporal Refinement (3D-PT-PST)}
\label{sec:proposed_method}

\subsection{Connectivity-Informed Power-Ladder Selection}

Let
$\mathcal{Q}=\{q_1,\ldots,q_M\}$, with
$q_1<\cdots<q_M$, denote a discrete set of candidate transmit powers
used during a pilot measurement stage. At candidate power $q_m$, pilot
packet detections are converted to
$b_{ij}^{(m)}\in\{0,1\}$, where $b_{ij}^{(m)}=1$ if the detected packet
fraction from tag $i$ to tag $j$ meets the pilot detection threshold.
The resulting matrix
$\mathbf{B}^{(m)}=[b_{ij}^{(m)}]$ is the adjacency matrix of the binary pilot graph at $q_m$.

The reciprocal-edge density of this graph is
\begin{equation}
\delta_m
=
\frac{2}{n(n-1)}
\sum_{i<j}
b_{ij}^{(m)}b_{ji}^{(m)},
\label{eq:pilot_density}
\end{equation}
i.e., the fraction of tag pairs with successful detections in both directions.

Let $\boldsymbol{\delta}^{\star}
=[\delta_1^{\star},\ldots,\delta_K^{\star}]$ denote fixed target
reciprocal-edge densities for progressively denser graphs. For each target, 3D-PT-PS selects the candidate power whose measured $\delta_m$ is closest to the target, subject to increasing transmit-power order. At the highest level, a connected reciprocal graph is preferred when available. The targets are fixed a priori without tag coordinates, ranges, RSSI, or localization error.

\subsection{3D-PT-PS Ordinal Reconstruction}

3D-PT-PS applies the multi-level reachability scoring and
reliability-weighted ordinal reconstruction of 3D-PT
\cite{moeini2026protraction3d}. The selected levels produce scores $s_{ij}$, reliabilities $\kappa_{ij}$, and weighted local and global ordinal relations optimized using $\mathcal{L}_{l}$ and $\mathcal{L}_{g}$. The resulting estimate initializes 3D-PT-PST.

\subsection{Temporal Refinement}

3D-PT-PST uses each tag's reader-recorded deposition timestamp
$\tau_i$ as a weak, non-metric cue. Tags are ordered as
$\tau_{\pi_1}\leq\cdots\leq\tau_{\pi_n}$, with
$\Delta\tau=\max_i\tau_i-\min_i\tau_i$; refinement is skipped when
$\Delta\tau$ is below the timestamp resolution. For
$\alpha_f\in\mathcal{A}$, let
$q_f=\max(1,\operatorname{round}(\alpha_f n))$ and consider
$(i,j)=(\pi_r,\pi_{r+q_f})$. The pair is retained when
$g_{ij}=(\tau_j-\tau_i)/\Delta\tau\geq g_{\min}$ and provides the soft
relation $z_i<z_j$, where $z_i$ is the coordinate along the inferred
deposition axis.

Because settlement, mixing, and rehandling can violate deposition
order, these relations are weighted rather than imposed as hard
constraints. Let
$\chi_i=(\max_j\tau_j-\tau_i)/\Delta\tau\in[0,1]$
denote the normalized residence age, with larger $\chi_i$ indicating
longer residence. Temporal reliability, margin, and unnormalized weight
are
\begin{equation}
\begin{aligned}
r_i^t &= \max(r_{\min},e^{-\beta\chi_i}), \qquad
\mu_{ij}^t = \mu_0+\mu_1g_{ij},\\
\widetilde w_{ij}^t
&= g_{ij}^{\gamma}\sqrt{r_i^t r_j^t}.
\end{aligned}
\label{eq:temporal_weights}
\end{equation}

At most $c_t n$ highest-weight relations are retained as
$\mathcal{E}$, with
$w_{ij}^t=\widetilde w_{ij}^t/\overline{w}^t$, where
$\overline{w}^t=|\mathcal{E}|^{-1}\sum_{(a,b)\in\mathcal{E}}
\widetilde w_{ab}^t$. The temporal loss is
\begin{equation}
\begin{aligned}
\mathcal{L}_{t}
&=
\frac{1}{|\mathcal{E}|}
\sum_{(i,j)\in\mathcal{E}}
w_{ij}^{t}\\
&\quad\times
\rho\!\left(
\frac{z_i-z_j+\mu_{ij}^{t}}{T_t}
\right).
\end{aligned}
\label{eq:temporal_loss}
\end{equation}
where $T_t>0$ is the temporal smoothing parameter and
$\rho(q)=c_{\rho}\log[1+\log(1+e^q)/c_{\rho}]$ is the 3D-PT
robust penalty, with $c_{\rho}=2.5$
\cite{moeini2026protraction3d}.

The deposition axis is inferred from the 3D-PT-PS embedding as the direction most correlated with the standardized entry timestamps.
Let $\eta_t$ denote the weighted fraction of retained temporal
relations satisfied by the radio-only 3D-PT-PS embedding. The effective
temporal weight is
\begin{equation}
\lambda_t^{\mathrm{eff}}
=
\lambda_t\!\left[
m_{\min}+(1-m_{\min})
\left[\frac{\eta_t-\eta_0}{\eta_1-\eta_0}\right]_0^1
\right],
\label{eq:temporal_adaptation}
\end{equation}
where $[x]_0^1=\min(1,\max(0,x))$. Thus, timestamp information is
down-weighted when its ordering disagrees with the radio-derived
geometry.

\subsection{3D Embedding Optimization}

The scale normalization and radio-only optimization follow 3D-PT \cite{moeini2026protraction3d}. 3D-PT-PST augments the 3D-PT-PS objective with the confidence-adaptive temporal loss:
\begin{equation}
\widehat{\mathbf{X}}
=
\arg\min_{\mathbf{X}}
\mathcal{L}_{l}
+\lambda_g\mathcal{L}_{g}
+\lambda_t^{\mathrm{eff}}\mathcal{L}_{t}
+\lambda_s\mathcal{L}_{s},
\label{eq:objective}
\end{equation}
where $\mathcal{L}_{s}$ is the scale-normalization loss inherited from
3D-PT, $\lambda_g$ and $\lambda_s$ are the global-order and
scale-normalization weights, respectively, and
$\lambda_t^{\mathrm{eff}}$ is the confidence-adaptive temporal weight
described above. For 3D-PT-PS, $\lambda_t^{\mathrm{eff}}=0$. For 3D-PT-PST, the 3D-PT-PS estimate provides the primary warm start. A localization is accepted only when the radio observations contain sufficient ordinal diversity, preventing the temporal cue alone from producing an apparently valid 3D reconstruction.

\section{Experimental Methodology}
\label{sec:experimental_methodology}

Two underground scales are considered: a $1000$~t storage pile, comparable to a reported ${\sim}1000$~t surge bin \cite{zhang2022carrapateena}, and a $5000$~t crusher-pocket stockpile; together they bracket reported $2000$--$2300$~t underground buffer capacities \cite{skawina2022comparison}. Piles are modeled as elliptical cones with angle of repose $\theta=37^{\circ}$ \cite{lindmark2018computational}, bulk density $\rho_b=2.0$~t/m$^3$ \cite{saadat2024flac3d,choudhary2019effect}, and nominal length-to-width ratios $1.45$ and $1.70$. For mass $m$,
\begin{equation}
\begin{aligned}
V &= \frac{m}{\rho_b}
   = \frac{\pi}{3}\,a\,b\,H,\\
b &= \frac{H}{\tan\theta},
\qquad
a = \left(\frac{L}{W}\right)b .
\end{aligned}
\label{eq:pile_geometry}
\end{equation}
where $L=2a$, $W=2b$, and $H$ are the pile dimensions. The nominal dimensions are approximately $22.0\times15.2\times5.7$~m and $41.8\times24.6\times9.3$~m, respectively. Let $r_0=L/W$ denote the nominal aspect ratio; each trial uses $r=r_0\max(0.75,1+0.08\xi)$, $\xi\sim\mathcal{N}(0,1)$. A Caterpillar~745 (${\sim}41$~t payload,
$5.90\times3.18\times1.40$~m bed) motivates the deposition scale
\cite{anas2025evaluation}. One tag opportunity is associated with each
nominal $40$~t dump, giving approximately $5/15/25$ tags for storage
and $25/75/125$ for the crusher pocket at $20/60/100\%$ tagging.

For each trial, an equivalent-volume 3D stockpile is generated using assumed bulk density, aspect ratio, and angle of repose. Tags are deposited sequentially according to truck-dump order with stochastic truck-scale dispersion. Dump order therefore creates only a statistical depth tendency rather than a deterministic vertical ordering. Simulator-only perturbations model lateral displacement, settlement, and occasional rehandling, which can further violate the initial deposition order. These parameters generate hidden ground truth and are never supplied to the localization methods.

Underground inter-dump intervals are drawn from a positive lognormal model with nominal mean $7$~min, coefficient of variation $0.30$, and minimum interval $1$~min. The recorded deposition-sequence timestamps are quantized to $1$-min resolution, and localization is evaluated $2$~min after the final dump.

The hardware model uses rugged sub-GHz mining tags operating as
peer-to-peer \emph{Wireless Positioning and Sensing Network (WPSN)}
nodes \cite{fcc2021muckpuck,moeini2024wireless}. Communication is modeled as proprietary LoRa-like signaling at $915$~MHz
\cite{moeini2025inventory}.

The simulated bulk-material path loss at separation $d$ is modeled as
\begin{equation}
PL(d)
=
PL_0
+
10\eta\log_{10}\!\left(\frac{d}{d_0}\right)
+
\alpha_b d
+
M_I,
\label{eq:propagation_model}
\end{equation}
Here $PL_0$ is the free-space reference loss at $d_0=1$~m, $\eta$ is the path-loss exponent, $\alpha_b$ is the bulk-material attenuation coefficient, and $M_I$ is an interference margin. For the nominal-ore model, $\eta=2$, $\alpha_b=1.5$~dB/m, and $M_I=2$~dB. Pair-shadowing, directional, and packet-level standard deviations are $6$, $2$, and $3$~dB, respectively, and receiver sensitivity is $-132$~dBm. These quantities generate hidden binary packet outcomes and are never supplied to localization.

Fixed entrance/exit event readers provide tag identity and minute-resolution event timestamps only; their spatial coordinates are never supplied to localization and they therefore do not act as localization anchors. The online temporal formulation uses entry  timestamps and does not use future exit information. Localization receives only binary packet-detection outcomes and event timestamps; RSSI, true inter-tag distances, pile dimensions, material parameters, truck dimensions, reader coordinates, and ground-truth positions are excluded.

For the reported experiments, the temporal design parameters are $\mathcal{A}=\{0.10,0.22,0.40,0.65\}$,
$g_{\min}=0.06$, $r_{\min}=0.35$, $\beta=1$,
$\mu_0=0.035$, $\mu_1=0.16$, $\gamma=1.20$,
$c_t=5$, and $T_t=0.20$. The base temporal coefficient is
$\lambda_t=0.30$, with $m_{\min}=0.15$, $\eta_0=0.50$, and
$\eta_1=0.78$. For $K=4$, fixed reciprocal-density targets
$\boldsymbol{\delta}^{\star}=[0.10,0.30,0.55,0.80]$ define the desired
graph regimes, but the realized four-power ladder is selected independently from each trial's binary pilot. The pilot candidate set is
$\mathcal{Q}=\{-80,-75,\ldots,20\}$~dBm. At each candidate power, eight pilot packet opportunities are used as a fixed design choice, and a directed detection is declared when the detected packet fraction is at least $0.80$. Sensitivity to pilot length is not investigated in the present study. MDS-Cal/CATL-Cal select the connected candidate closest to density $0.35$ when possible; MDS Max/CATL-Max use $20$~dBm.

Two experiments are performed. The primary experiment uses the
nominal-ore model with $K=4$ and evaluates six methods: MDS-Max and
CATL-Max, which denote MDS-MAP(P) and CATL operating with
maximum-power connectivity, respectively; MDS-Cal and CATL-Cal,
which denote MDS-MAP(P) and CATL operating at a single transmit-power
level calibrated from the binary pilot graph, respectively; 3D-PT-PS
using multi-level reachability; and 3D-PT-PST additionally using
entry timestamps.
Each 3D-PT-PS/3D-PT-PST level uses 20 localization packet opportunities, giving 80 per tag at $K=4$. Each single-power baseline receives the same 80-packet localization budget at its respective power. A total of 250 independent trials are performed per scenario and tag fraction, with geometry and latent-channel realizations paired across methods.

For $K\in\{1,2,4,6,8\}$, a separate fixed-budget ablation is performed in the $5000$~t crusher-pocket scenario for 25, 75, and 125 tags. The localization budget is fixed at 72 packet opportunities per tag, giving $72/K$ packets per power level. The binary pilot uses the same measurement overhead for every $K$ and is separate from this localization budget. The $K$-sweep uses 25 trials per condition, with geometry and latent-channel realizations paired across $K$.

After similarity alignment, let $\widetilde{\mathbf{x}}_i$ denote the aligned estimate and define the true network diameter as $D_{\max}=\max_{i<j}\|\mathbf{x}_i-\mathbf{x}_j\|_2$.

The normalized root-mean-square error (NRMSE) is
\begin{equation}
\mathrm{NRMSE}
=
\frac{100}{D_{\max}}
\sqrt{
\frac{1}{n}
\sum_{i=1}^{n}
\left\|
\widetilde{\mathbf{x}}_i-\mathbf{x}_i
\right\|_2^2
}.
\label{eq:nrmse}
\end{equation}

Performance is evaluated using NRMSE, 3D and axis-specific root-mean-square error (RMSE), pairwise stress, localization success rate, and runtime. Mean NRMSE and axis-specific RMSE are summarized with $95\%$ bootstrap confidence intervals using 5000 resamples in the primary experiment and 1000 in the $K$-sweep. Two-sided Wilcoxon signed-rank tests are applied to paired per-trial NRMSE values; paired mean differences are additionally summarized with $95\%$ bootstrap confidence intervals.

All simulations and localization methods were implemented in MATLAB R2024b and executed on an ASUS ROG Zephyrus G16 equipped with an Intel Core Ultra 9 185H processor (2.50~GHz) and 32~GB of RAM.

\section{Results and Discussion}
\label{sec:results}

Table~\ref{tab:main_results} summarizes the primary $K=4$ experiment
using the calibrated single-power baselines and both 3D-PT variants.
Across all six scenario--density conditions, 3D-PT-PST achieves
the lowest mean NRMSE, while 3D-PT-PS outperforms the calibrated
baselines.

\begin{table}[t]
\centering
\caption{Mean localization NRMSE (\%) with 95\% bootstrap confidence
intervals for $K=4$. PT-PS and PT-PST denote 3D-PT-PS and
3D-PT-PST, respectively.}
\label{tab:main_results}
\setlength{\tabcolsep}{2.3pt}
\renewcommand{\arraystretch}{1.30}
\footnotesize
\begin{tabular}{lccccc}
\toprule
Scenario & $n$ & MDS-Cal & CATL-Cal & PT-PS & PT-PST \\
\midrule
Storage & 5
& $22.15_{21.38}^{23.00}$
& $34.67_{33.78}^{35.55}$
& $17.88_{17.17}^{18.57}$
& $\mathbf{17.46_{16.82}^{18.13}}$ \\

Storage & 15
& $20.49_{20.12}^{20.87}$
& $24.14_{23.47}^{24.81}$
& $17.67_{17.34}^{18.01}$
& $\mathbf{15.67_{15.25}^{16.10}}$ \\

Storage & 25
& $18.43_{18.16}^{18.72}$
& $22.01_{21.50}^{22.53}$
& $14.96_{14.71}^{15.21}$
& $\mathbf{11.68_{11.42}^{11.93}}$ \\

Crusher & 25
& $16.01_{15.75}^{16.26}$
& $20.29_{19.66}^{20.94}$
& $11.63_{11.45}^{11.83}$
& $\mathbf{8.98_{8.81}^{9.15}}$ \\

Crusher & 75
& $12.01_{11.88}^{12.14}$
& $19.25_{18.79}^{19.72}$
& $7.87_{7.73}^{8.01}$
& $\mathbf{6.20_{6.13}^{6.27}}$ \\

Crusher & 125
& $10.91_{10.81}^{11.01}$
& $20.26_{19.85}^{20.69}$
& $7.08_{6.95}^{7.22}$
& $\mathbf{6.12_{6.06}^{6.18}}$ \\
\bottomrule
\end{tabular}
\end{table}
\begin{figure}[t]
    \centering
    \begin{minipage}[t]{0.49\columnwidth}
        \centering
        \includegraphics[width=\linewidth]{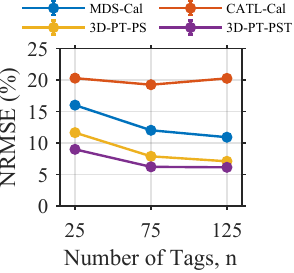}
        \vspace{-1mm}
        \centerline{\small (a)}
    \end{minipage}
    \hfill
    \begin{minipage}[t]{0.49\columnwidth}
        \centering
        \includegraphics[width=\linewidth]{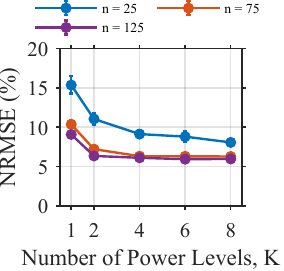}
        \vspace{-1mm}
        \centerline{\small (b)}
    \end{minipage}
    \caption{Localization performance in the $5000$~t crusher-pocket scenario:
    (a) mean NRMSE versus tag count at $K=4$;
    (b) 3D-PT-PST NRMSE versus $K$ under a fixed 72-packet localization budget. Error bars show $95\%$ bootstrap confidence intervals.}
    \vspace{-4mm}
    \label{fig:crusher_results}
\end{figure}

Figure~\ref{fig:crusher_results}(a) shows improved accuracy with tag
density: 3D-PT-PST decreases from $8.98\%$ NRMSE at 25 tags to
$6.20\%$ at 75 tags and $6.12\%$ at 125 tags. The small 75-to-125-tag
change indicates diminishing returns, while the five-tag storage case
($17.46\%$) illustrates the difficulty of sparse 3D reconstruction.

Figure~\ref{fig:crusher_results}(b) confirms the benefit of multi-level reachability under a fixed packet budget. At 125 tags, NRMSE falls from $9.08\%$ at $K=1$ to $6.11\%$ at $K=4$ and $5.96\%$ at $K=8$. Most improvement is therefore obtained by $K=4$, with larger gains from additional levels mainly in the sparse case. This supports $K=4$ as the primary accuracy--measurement-complexity operating point.

At 125 tags, 3D-PT-PST reduces mean NRMSE from $7.08\%$ for 3D-PT-PS to $6.12\%$ ($13.5\%$ relative), with a paired mean reduction of $0.96$ percentage points (95\% CI: $0.82$--$1.11$; Wilcoxon $p<0.001$). Relative to MDS-Cal and CATL-Cal, 3D-PT-PST gives paired reductions of $43.6\%$ and $69.1\%$, respectively, with $p<0.001$. The temporal variant primarily improves the deposition axis: its $x/y/z$ RMSE values are $0.93/0.93/1.76$~m versus $0.96/0.95/2.15$~m for 3D-PT-PS, yielding $2.20$~m 3D RMSE.

Both PT variants succeed in 1499 of 1500 primary trials ($99.93\%$);
success requires sufficient radio-derived ordinal information, a finite
3D estimate, and valid similarity alignment. At 125 tags, mean runtimes are $71.0$~s for 3D-PT-PS and $57.5$~s for 3D-PT-PST, versus $2.09$~s for MDS-Cal, favoring periodic reconstruction over high-rate tracking.

\section{Conclusion}
\label{sec:conclusion}

This paper introduced 3D-PT-PS with connectivity-informed binary power-ladder selection and 3D-PT-PST with confidence-adaptive temporal refinement \cite{moeini2026protraction3d}. Under a fixed packet budget, most of the multi-level gain is obtained by $K=4$, while entry timestamps further improve the deposition-axis estimate. In the $5000$-t, 125-tag crusher-pocket condition, 3D-PT-PST achieves $6.12\%$ mean NRMSE and $2.20$~m 3D RMSE without RSSI, measured ranges, surveyed anchors, pile dimensions, or known material parameters. Future work will validate 3D-PT-PST using buried-tag measurements.

\vfill\pagebreak

\bibliographystyle{IEEEbib}
\bibliography{refs}

@article{geok2021review,
  author  = {{Geok, T. K. and Aung, K. Z. and Aung, M. S. and Soe, M. T. and Abdaziz, A. and Liew, C. P. and Hossain, F. and Tso, C. P. and Yong, W. H.}},
  title   = {{Review of Indoor Positioning: Radio Wave Technology}},
  journal = {{Applied Sciences}},
  volume  = {11},
  pages   = {279},
  year    = {2021},
  doi     = {10.3390/app11010279}
}

@article{theissen2023usecase,
  author  = {{Theissen, M. and Kern, L. and Hartmann, T. and Clausen, E.}},
  title   = {{Use-Case-Oriented Evaluation of Wireless Communication Technologies for Advanced Underground Mining Operations}},
  journal = {{Sensors}},
  volume  = {23},
  pages   = {3537},
  year    = {2023},
  doi     = {10.3390/s23073537}
}

@article{both2021applied,
  author  = {{Both, C. and Dimitrakopoulos, R.}},
  title   = {{Applied Machine Learning for Geometallurgical Throughput Prediction—A Case Study Using Production Data at the Tropicana Gold Mining Complex}},
  journal = {{Minerals}},
  volume  = {11},
  pages   = {1257},
  year    = {2021},
  doi     = {10.3390/min11111257}
}

@article{moeini2026protraction,
  author  = {{Moeini, P. and Coates, M.}},
  title   = {{Protraction Technique: Anchor-Free Range-Free Localization From Multi-Level Binary Reachability in Wireless Sensor Networks}},
  journal = {{IEEE Access}},
  volume  = {14},
  pages   = {131894--131910},
  year    = {2026},
  doi     = {10.1109/ACCESS.2026.3726898}
}

@inproceedings{shang2004improved,
  author    = {{Shang, Y. and Ruml, W.}},
  title     = {{Improved MDS-Based Localization}},
  booktitle = {{IEEE INFOCOM 2004}},
  volume    = {4},
  pages     = {2640--2651},
  year      = {2004},
  doi       = {10.1109/INFCOM.2004.1354683}
}

@article{tan2013catl,
  author  = {Tan, Guang and Jiang, Hongbo and Zhang, Shengkai and Yin, Zhimeng and Kermarrec, Anne-Marie},
  title   = {Connectivity-based and anchor-free localization in large-scale {2D/3D} sensor networks},
  journal = {ACM Trans. Sensor Netw.},
  volume  = {10},
  number  = {1},
  pages   = {1--21},
  month   = sep,
  year    = {2013}
}

@inproceedings{zhang2022carrapateena,
  author    = {{Zhang, Q. and Hocking, R. J.}},
  title     = {{Carrapateena Block Cave Mine Design and Planning: Feasibility Study}},
  booktitle = {{Caving 2022: Proceedings of the Fifth International Conference on Block and Sublevel Caving}},
  editor    = {{Potvin, Y.}},
  publisher = {{Australian Centre for Geomechanics}},
  address   = {Perth, Australia},
  pages     = {291--302},
  year      = {2022},
  doi       = {10.36487/ACG_repo/2205_19}
}

@article{skawina2022comparison,
  author  = {{Skawina, B. and Salama, A. and Gunillasson, J. and Strömsten, M. and Wettainen, T.}},
  title   = {{Comparison of Productivity When Running Filled, Near-Empty, or Flow-Through Orepass Using Discrete Event Simulation}},
  journal = {{Mining}},
  volume  = {2},
  pages   = {186--196},
  year    = {2022},
  doi     = {10.3390/mining2020011}
}

@article{lindmark2018computational,
  author  = {{Lindmark, D. M. and Servin, M.}},
  title   = {{Computational Exploration of Robotic Rock Loading}},
  journal = {{Robotics and Autonomous Systems}},
  volume  = {106},
  pages   = {117--129},
  year    = {2018},
  issn    = {0921-8890},
  doi     = {10.1016/j.robot.2018.04.010}
}

@article{choudhary2019effect,
  author  = {{Choudhary, B.}},
  title   = {{Effect of Blast Induced Rock Fragmentation and Muckpile Angle on Excavator Performance in Surface Mines}},
  journal = {{Mining of Mineral Deposits}},
  volume  = {13},
  pages   = {119--126},
  year    = {2019},
  doi     = {10.33271/mining13.03.119}
}

@article{saadat2024flac3d,
  author  = {{Saadat, M. and Khishvand, M. and Seccombe, A.}},
  title   = {{FLAC3D Simulation of Caving Mechanism and Strata Fracture Response in Underground Mining}},
  journal = {{Mining}},
  volume  = {4},
  pages   = {818--840},
  year    = {2024},
  doi     = {10.3390/mining4040046}
}

@article{anas2025evaluation,
  author  = {{Anas, A. and S, W. and Tui, R. and Amalia, R.}},
  title   = {{Evaluation of Articulated CAT 745 Fuel Ratio Based on Haul Road Geometry at PT Hillconjaya Sakti Site PT Adhi Kartiko Pratama Tbk}},
  journal = {{EPI International Journal of Engineering}},
  volume  = {8},
  pages   = {15--25},
  year    = {2025},
  doi     = {10.25042/epi-ije.022025.03}
}

@misc{fcc2021muckpuck,
  author       = {{Federal Communications Commission}},
  title        = {{Grant of Equipment Authorization, FCC ID 2AYG6MP2001 (Peytec Inc., MuckPuck Model RMTS-XT-MP-004)}},
  howpublished = {{FCC Office of Engineering and Technology, Equipment Authorization System}},
  year         = {2021},
  note         = {{ISED Canada Certification 26799-MP2001}}
}

@misc{moeini2025inventory,
  author       = {{Moeini, P. and Sinha, A. and Ahmari, A. and Hogiu, S. and Guergachi, S. and Sadrmahdavi, O.}},
  title        = {{Inventory System for Mines}},
  howpublished = {{U.S. Patent Application Publication US 2025/0274722 A1}},
  year         = {2025},
  note         = {{Assignee: Peytec Inc.; filed Apr. 11, 2022; published Aug. 28, 2025}}
}

@misc{moeini2024wireless,
  author       = {{Moeini, P. and Murari, R. and Mohammadi, F. and Ahmari, A. and Al-Obaidy, F. and Asad, A. and Fernando, D. and Fernando, X. N. and Guergachi, S. and Jaseemuddin, M. and Li, X.}},
  title        = {{Wireless Positioning Systems}},
  howpublished = {{U.S. Patent 11,914,856 B2}},
  year         = {2024},
  note         = {{Assignee: 14957539 Canada Inc.; filed Jan. 8, 2021; granted Feb. 27, 2024}}
}

@misc{moeini2026protraction3d,
  author = {{Moeini, P. and Coates, M.}},
  title  = {{3D Protraction Technique for Anchor-Free, Range-Free Wireless Localization in Industrial Warehouses via Multi-Level Binary Reachability}},
  note   = {{arXiv preprint arXiv:2609.09330}},
  year   = {2026},
  doi    = {10.48550/arXiv.2609.09330}
}

@article{young2021modelling,
  author  = {{Young, A. and Rogers, W. P.}},
  title   = {{Modelling Large Heaped Fill Stockpiles Using FMS Data}},
  journal = {{Minerals}},
  volume  = {11},
  pages   = {636},
  year    = {2021},
  doi     = {10.3390/min11060636}
}

@article{marinin2021assessing,
  author  = {{Marinin, M. and Marinina, O. and Wolniak, R.}},
  title   = {{Assessing of Losses and Dilution Impact on the Cost Chain: Case Study of Gold Ore Deposits}},
  journal = {{Sustainability}},
  volume  = {13},
  pages   = {3830},
  year    = {2021},
  doi     = {10.3390/su13073830}
}

@article{vanduijvenbode2020performance,
  author  = {{van Duijvenbode, J. R. and Buxton, M. W. N. and Shishvan, M. S.}},
  title   = {{Performance Improvements during Mineral Processing Using Material Fingerprints Derived from Machine Learning---A Conceptual Framework}},
  journal = {{Minerals}},
  volume  = {10},
  pages   = {366},
  year    = {2020},
  doi     = {10.3390/min10040366}
}

\end{document}